\documentstyle[12pt]{article}
\begin{document}
\begin{titlepage}
\rightline {Si-95-11 \  \  \  \   }
\rightline {arch-ive/9511155 \  \  \  \   }

\vskip 2truecm
\centerline{\Large\bf A Generalization of the Bargmann-Fock}
\centerline{\Large\bf Representation to Supersymmetry}
\centerline{\Large \bf  by Holomorphic Differential Geometry}

\vskip 1.0truecm

\centerline{PACS Nos.: 11.30, 3.65, 2.40}

\vskip 2.0truecm

\centerline{\bf H.-P. Thienel}

\vskip 0.7truecm

\centerline{Universit\"at Siegen, Fachbereich Physik,
D-57068 Siegen, Germany}
\centerline{e-mail: thienel@sicip1.physik.uni-siegen.de}

\vskip 2.0truecm

\abstract

\noindent
In the Bargmann-Fock representation the coordinates $z^i$ act as
bosonic creation operators while the
partial derivatives $\partial_{z^j}$ act as
annihilation operators on holomorphic $0$-forms as states of a
$D$-dimensional bosonic oscillator.
Considering also $p$-forms and further geometrical objects as
the exterior derivative and Lie derivatives on a
holomorphic ${\bf C}^D$, we end up with an analogous representation
for the $D$-dimensional supersymmetric oscillator. In particular,
the supersymmetry  multiplet structure of the Hilbert space
corresponds to the cohomology of the exterior derivative.
In addition, a 1-complex parameter group emerges naturally and
contains both time evolution and a homotopy related to cohomology.
Emphasis is on calculus.

\end{titlepage}

\section{Introduction}

  The
  conventional
   Bargmann-Fock representation displays
the D-\-dimensional bosonic harmonic oscillator by using
holomorphic $0$-forms on a manifold ${\bf C}^D$ to represent states
in a Hilbert space.
The bosonic creation and annihilation operators are
represented by $z^i, \partial_{z^j}$ respectively \cite{foc,bar,bar1}.
Thus, the Bargmann-Fock representation is a local geometrical concept.
We extend this idea, including also holomorphic $p$-forms and
considering  geometrical operations on these, to  find that
the  $D$-dimens\-ional supersymmetric (SUSY) oscillator
\cite{nic,wit,wit1,rit,gen}
is realized in every  detail.

One way to develop the formalism
would be to reformulate
the local differential geometry of
the full ${\bf C}^D$ and impose the restriction to
holomorphic quantities afterwards, since the standard geometrical
apparatus is more familiar for the more general case \cite{cho,wel}.
Instead, in the second section,
we will develop a calculus for a
purely  holomorphic differential
geometry and,
by immediate interpretation, built up
the corresponding physical system simultaneously.
The only structure that is not really
geometric on a holomorphic manifold but
an additional ingredient is the scalar product, of which
we give an alternative definition not involving any integral
which would exceed
the concept of holomorphic geometry.
We will see that supersymmetry
(SUSY) is supplied by the operator $\partial$.
The Hamiltonian is a Lie derivative corresponding
to a 1-complex-parameter group
that can be split by holomorphicity into two equivalent
1-parameter groups. One can be identified with
evolution in a Hilbert space. The other supplies
a homotopy that
occurs in the
proof of Poincar\'e's lemma \cite{cho}.
We end this section
 with  a brief discussion of the eigenstates of two Lie derivatives,
the first one being the Hamiltonian and the second one
yielding coherent states.

The concluding remarks of the third section
will first comment on the
underlying supergroup structure.
Secondly,
for completeness, we will give an
integral expression for the scalar product
accomodating also $p$-forms. This will
relate our representation, which
dispenses with
square integrable functions, to the coherent
state representation and  other familiar
representations, where square integrable functions
are employed.
Finally, we give a prescription that would give us the
physical states represented by holomorphic forms,
if we had started with the
geometry of the full ${\bf C}^D$.

In the following,
 all commutators are graded ones, i.e. if both
entries have odd form degree, we have an anti-commutator, otherwise we have
a commutator. The type of the commutator is indicated by a subscript for the
convenience of the reader. The same  indices in upper and lower
position indicate a sum from $1$ to $D$.

\section{Holomorphic geometry and the SUSY oscillator}

Consider a manifold ${\bf R}^{2D}$ and choose a global parametrization
$x^j,y^j , \quad j=1,2,...,D$, where both $x^j$ and $y^j$ take values
in ${\bf R}$.
We combine pairs of real coordinates into
$z^j=x^j+iy^j $ taking values in ${\bf C}$, such that our manifold is now
${\bf C}^D$. Furthermore, we demand that any function ${\bf C}^D \to {\bf C}$
be of the
form
 \begin{equation}
 f(z^i)=f_0+f_k z^k+ f_{kl}z^k z^l+..., \qquad f_{kl...}
={\rm const.} \in {\bf C},
\label{1}
\end{equation}
i.e. holomorphic around the origin $z^1=z^2=...=z^D=0$.
The series eq.(\ref{1})
has to be convergent on all of ${\bf C}^D$ and we call a manifold, where
such functions live  ${\bf C}^D_{\rm h}$ (holomorphic ${\bf C}^D$ ).
In fact, $f(z^i)< const. \exp(-\sum^D_{k=1}(z^k)^2/2)$
\cite{bar1}, in order to
yield normalizable states.

Besides the functions, it is natural to have holomorphic vector fields
on our manifold. Since we work in a fixed frame, there is a canonical
decomposition
\begin{equation}
v=v^i \partial_{z^j}\in T{\bf C}^D_{\rm h},\label{2}
\end{equation}
where $v^i(z^i)$ are holomorphic functions as in eq.(\ref{1})
and $\partial_{z^j} ={1 \over 2} (\partial_{x^j}-i \partial_{y^j})$ are the
holomorphic basis vectors $\in T{\bf C}^D_{\rm h}$.
Among the coordinates $z^i$ and the basis vectors $\partial_{z^j}$
the following commutation relations hold due to
$  \partial_{z^j} z^i=\delta_j^i$
\begin{equation}
[z^i,z^j]_-=[\partial_{z^i},\partial_{z^j}]_-=0,
\qquad [\partial_{z^i},z^j]_-=\delta^j_i,
\label{3}
\end{equation}
which is the algebra of bosonic creators $z^i$ and
annihilators $\partial_{z^j}$
operating on functions eq.(\ref{1}).

{}From this point of view, we can apply $z^i$ to a ''vacuum'' $1$
(from the left)
in order to get eq.(\ref{1}), which represents a general state
in a bosonic Fock space. Along with the tangent space
$T{\bf C}^D_{\rm h}$ of the holomorphic vectors  $v$,
the dual cotangent space
$T^*{\bf C}^D_{\rm h}$, containing the holomorphic 1-forms $F^{(1)}$
that provide linear maps of the holomorphic vectors to
${\bf C}$, is a natural geometrical structure. Again, there
is a canonical decomposition $F_j(z^i)dz^j$
with $F_j$ holomorphic as in eq.(\ref{1}) and $dz^j=dx^j+idy^j$.
A holomorphic $p$-form may be written as
\begin{equation}
 F^{(p)}(z^i,dz^j)
 =F_{i_k,...,k_p}( z^i)dz^{k_1}....dz^{k_p}\in \Lambda^p {\bf C}^D_{\rm h},
\label{4}
 \end{equation}
with the factors $F_{k_1,...,k_p}(z^i)$
as in eq.(\ref{1}). Observe that $0\le p\le D$, although the dimension
of ${\bf C}^D_{\rm h}$ is $2D$.
Finally, holomorphic forms are  (finite) power series in the
 $dz^j$
\begin{equation}
 \Psi ( z^i, dz^j)=F_0( z^i)+F_k( z^i) dz^k
+ F_{kl}( z^i)dz^k dz^l+... \quad \in \Lambda {\bf C}^D_{\rm h}
=\bigoplus_{p=0}^D \Lambda^p {\bf C}^D_{\rm h},
\label{5}
\end{equation}
spanning the holomorphic exterior algebra.

On holomorphic forms the interior derivative
is a natural geometrical operation that
maps $\Lambda^p {\bf C}^D_{\rm h}$ to $ \Lambda^{p-1} {\bf C}^D_{\rm h}$
by contraction with a holomorphic vector $v$.
As used in \cite{thi}
an interior derivative on a real manifold induced by a real vector $u^i
\partial_{x^i}$ can be written as
\begin{equation}
u^i \partial_{dx^i}:=u^i \frac{\partial}{\partial dx^i}\equiv i_u ,
\label{6}
\end{equation}
where $ \partial / \partial_{dx^i}$ is a Grassmann left derivative
with respect
to the Grassmann number $dx^i$. The duality of frame and coframe
is expressed by $ \partial dx^j / \partial dx^i =\delta^j_i$.
Accordingly, a vector $v=v^k(z^i)\partial_{z^k}=
{1 \over 2}v^k(z^i)\partial_{x^k}-{i \over 2}v^k(z^i)\partial_{y^k}$
induces an interior derivative
\begin{equation}
v^i \partial_{dz^i}:=v^i \frac{\partial}{\partial dz^i}\equiv i_v ,
\label{7}
\end{equation}
where we have defined a new Grassmann left derivative
$\partial_{dz^j} ={1 \over 2} (\partial_{dx^j}-i \partial_{dy^j}).$
(On ${\bf C}^D$, a  general vector $w=w_z^k(z^i,\bar{z}^j)\partial_{z^k}
+w_{\bar{z}}^k(z^i,\bar{z}^j)\partial_{\bar{z}^k}$ induces the interior
derivative $i_w\equiv w_z^k\partial_{dz^k}
+w_{\bar{z}}^k\partial_{d\bar{z}^k})$.
The duality of holomorphic frame $\partial_{z^i}$  and
holomorphic coframe $dz^j$
 is expressed by $ \partial dz^j / \partial dz^i=\delta^j_i$.
This gives rise to
\begin{equation}
[dz^i,dz^j]_+=[\partial_{dz^i},\partial_{dz^j}]_+=0,
\qquad [\partial_{dz^i},dz^j]_+=\delta^j_i,
\label{8}
\end{equation}
which is the algebra of fermionic creation operators $dz^j$ and
annihilation operators $\partial_{dz^j}$. The commutators
of mixed bosonic and fermionic entries are
\begin{equation}
[z^i,dz^j]_-=[z^i,\partial_{dz^j}]_-=[\partial_{z^i},dz^j]_-=
[\partial_{z^i},\partial_{dz^j}]_-=0.
\label{9}
\end{equation}
Thus, a form eq.(\ref{4}) constitutes a general
state vector
of a Fock space with $D$ bosonic and $D$ fermionic degrees of freedom

The next task is to introduce a scalar product that makes the
exterior algebra a Hilbert space. We start by defining the adjoint operation.
Inspired by the algebraic properties of the elementary operators
$z^i,dz^j,\partial_{z^k},\partial_{dz^l}$, we define
their  adjoints by
\begin{equation}
 (z^i)^+:=\partial_{z^i},\qquad
 (dz^j)^+:=\partial_{dz^j},
\label{10}
\end{equation}
with the rules
$ (A+B)^+=A^+ +B^+$, $ (AB)^+=B^+A^+$ $(A^+)^+=A$ and $c^+=c^*$, if
$c=const.  \in {\bf C}$. While a state vector $\Psi(z^i,dz^j)$
    is a power series in $z^i,dz^j$,
its dual  $\Psi^+( \partial_{dz^l} , \partial_{z^k})$
    is a power series in $\partial_{z^k},\partial_{dz^l}$.
The prescription for the scalar product of two state vectors $\Psi$
and $\Xi$    from the
exterior algebra is the following
\begin{equation}
\langle \Psi(z^i,dz^j)|\Xi(z^k,dz^l)\rangle :=\Psi^+
(\partial_{dz^j} , \partial_{z^i})\Xi(z^k,dz^l)|_{z^1=z^2=...z^D=dz^1=dz^2=
...=dz^D=0},
\label{11}
\end{equation}
i.e. perform all the derivations in $\Psi^+$   on $\Xi$   and put
the remaining factors $z^i,dz^j$     to zero.

Up to now, we implicitely used the exterior derivative $d$
that maps bosonic $z^j$ to fermionic $dz^j$, which may be decomposed
as $d=\partial+\bar{\partial}=dz^i \partial_{z^i}
+d\bar{z}^i \partial_{\bar{z}^i}$. Since $\bar{\partial}=0$
on ${\bf C}^D_{\rm h}$, $d$ reduces to $\partial$.
 It is nilpotent $\partial^2=0$. Its adjoint is
$\partial^+=z^i \partial_{dz^i}$
   mapping fermionic $dz^j$  to
bosonic $z^j$, being also  nilpotent $(\partial^+)^2=0$.
$\partial^+$ is an interior derivative with respect to the vector field $z^i
\partial_{z^i} $.

The final
prominent geometrical object that we consider is the holomorphic
Lie derivative with
respect to a vector field $v$, which is an anti-commutator
\begin{equation} {\cal L}_v=[\partial, v^i \partial_{dz^i}]_+.
\label{12}
\end{equation}
(On ${\bf C}^D$, by definition $L_w=[\partial+ \bar{\partial},
w^i_z\partial_{dz^i}
+ w^i_{\bar{z}}\partial_{\bar{dz}^i}]_+$ corresponds
to a 1-complex-parameter group. There is a
decomposition $L_w= {\cal L}_w +\bar{\cal L}_w:=[\partial,w^i
\partial_{dz^i}+w^i_{\bar{z}}\partial_{d\bar{z}^i}]_+
+ [\bar{\partial},w^i_z
\partial_{dz^i}+w^i_{\bar{z}}
\partial_{d\bar{z}^i}]_+$, where
each ${\cal L}_w$ and $\bar{\cal L}_w $ corresponds to a
1-complex-parameter group.)
{}From eq.(\ref{12}), an important property follows immediately
\begin{equation}[ {\cal L}_v,\partial]_-=0.
\label{13}
\end{equation}

The Lie derivative is selfadjoint, if the interior derivative is
adjoint to $\partial$. This is true for
\begin{equation}
H:= {\cal L}_{z^i \partial_{z^i}}=[\partial,\partial^+]_+=
z^i \partial_{z^i}+dz^i\partial_{dz^i}
\label{14}
\end{equation}
The first term counts the powers in the coordinates of an expression, that
it is applied on. So we define the boson number operator
$N:=z^i\partial_{z^i}$,
of form degree $0$.
The second term
counts the form degree, if it is applied to a $p$-form.
Accordingly, we define the fermion number operator
$P:=dz^i\partial_{dz^i}$,
which also has form degree $0$.
Due to eq.(\ref{13}), $\partial$ and $\partial^+$ are conserved
\begin{equation}
[\partial,H]_-=[\partial^+,H]_-=0.
\label{15}
\end{equation}
Equations (\ref{14}) and (\ref{15}) represent the algebra of a
$D$-dimensional
SUSY oscillator with the SUSY
Hamiltonian $H$ and charges $\partial$ and $\partial^+$,
which can be combined
into two self-adjoint charges $Q_1=\partial+\partial^+$ and
$Q_2=i(\partial-\partial^+)$,
such that we have $N=2$ SUSY.

Locally a holomorphic Lie derivative is just a total derivative with
respect to a complex parameter $\theta=\tau+it$.
Since for our trivial topology local and global concepts coincide,
we have on the entire space
\begin{equation}
-{d \over d\theta} \Psi(\theta)=H\Psi(\theta)
\label{16}
\end{equation}
for any
\begin{equation}
\Psi(\theta)=\Psi( z^i(\theta), dz^j(\theta))
=\Psi( z^i e^{-\theta},dz^j e^{-\theta}  )
=e^{-\theta H}\Psi( z^i  , dz^j  )
\in \Lambda {\bf C}^D_{\rm h},
\label{17}
\end{equation}
where $z^i \equiv z^i(0)$ and $dz^j \equiv dz^j(0)$

Hence,  holomorphic forms are subject to an ''evolution''
in the parameter $ \theta$
 corresponding to a trivial
line bundle of charts:
${\bf C}(\theta) \times {\bf C}^D_{\rm h}(z^1,...,z^D)$.
$\theta$ parametrizes a sequence of charts
on ${\bf C}^D_{\rm h}$ each representing
the system at an ''instant'' $\theta$.
Although each chart is a Hilbert space, the
whole line bundle ${\bf C}(\theta) \times {\bf C}^D_{\rm h}(z)$,
 representing a complex  ''evolution'', is not a Hilbert space.

It is however possible to use inherent information to eliminate the
euclidean $\tau$ such that the remaining bundle   is a Hilbert space
as we are accustumed to.
Along with eq.(\ref{16})
\begin{equation}{d \over d \bar{\theta}} \Psi(\theta)=0
\label{18}
\end{equation}
also holds, since $\Psi(z^i e^{-\theta},dz^j e^{-\theta})$
is holomorphic in $\theta$.
We use eq.(\ref{18}) to eliminate $\tau$ from eq.(\ref{16})
\begin{equation}
i{d \over d t} \Psi(t)=H\Psi(t)
\label{19}
\end{equation}
This identifies $t$ as the time and the system evolves within
a bundle of charts
$U(1)(t)\times {\bf C}^D_{\rm h}(z^1, ...,z^D)$,
which is a Hilbert space as a whole and
the scalar product is preserved in t-evolution automatically.

We can conversely eliminate $t$ and we end up with
\begin{equation}
-{d \over d\tau} \Psi(\tau)=H\Psi(\tau)
\label{20}
\end{equation}
corresponding to an alternative bundle
${\bf R}_+(\tau)\times {\bf C}^D_{\rm h}(z^1,...,z^D).$
$\tau$-evolution preserves the scalar product due to $\tau^+=-\tau$,
 but the
bundle itself is not a Hilbert space, cf.
\cite{thi}.
Considering the limit $\tau \to \infty$ of eq.(\ref{17}),
 by use of eq.(\ref{14}),
we prove that any state with $p \ge1$
 is $\partial$-exact, if and only if  it is $\partial$-closed,
which is Poincar\'e's lemma on $C^D_{\rm h}$.
More precisely, we find that the only non-trivial cohomology class
contains the  constant numbers with  zero eigenvalue of $H$,
representing the normalized state $1$, which is  the only state
 left in the limit  $\tau \to \infty$.
It thus follows that, with exception of
 the SUSY singlet vacuum $1$, all
other states are paired by the operator $\partial$.
This conforms, of course, with the fact that the conditions
$\partial\Omega=0$, while $\Omega \ne \partial...$,
alternatively expressed by
$\partial\Omega=
\partial^+\Omega=0$, are  directly solved by $\Omega=const.$

We emphasize that eqns. (\ref{19}) and (\ref{20}) on ${\bf C}^D_{\rm h}$
are equivalent. The first one describes evolution on
a Hilbert space, the second one encodes the
cohomology of the underlying manifold
by supplying a homotopy of the manifold, which renders the non-trivial
cohomology classes in the  limit $\tau \to \infty$.
This makes transparent the intimate relation between the SUSY
and the time-evolution which is a characteristic feature
of any SUSY theory.

Since the holomorphic Lie derivative is an
operator of form degree zero, it  commutes
with the fermion number operator $P$, of which the eigenstates are
homogeneous
$p$-forms corresponding to the eigenvalue $p$. Hence the eigenstates
of a holomorphic
Lie derivative  can always be arranged to yield homogeneous $p$-forms.
The $\partial$-operator provides a pairing by mapping
 a non-closed $p$-form to an exact $p+1$-form, which both
span a 2-dimensional eigenspace corresponding to an eigenvalue of
${\cal L}_v$
(up to further degeneracy not related to SUSY).
Two Lie derivatives are particularly interesting:

i) the Hamiltonian $H$ yielding a complete set of
eigenstates given by  monomials
\begin{equation}
\Phi_E^{(p)}
={1 \over \sqrt{n_1!n_2!...n_D!}}
(z^1)^{n_1}(z^2)^{n_2}...(z^D)^{n_D}(dz^1)^{p_1}(dz^2)^{p_2}...
(dz^D)^{p_D} ;
\label{22}
\end{equation}
$$n_j=0,1,2,...; \quad p_j=0,1;\quad p=p_1+p_2+...+p_D;$$
corresponding to the energy value $E=n_1+n_2+...+n_D+p_1+p_2+...+p_D$ and
being orthonormal
\begin{equation}
\langle  \Phi_E^{(p)},
\Phi_{E'}^{(p')}\rangle =
\delta_{n_1 n'_1}\delta_{n_2 n'_2 }...\delta_{n_D n'_D}
\delta_{p_1 p'_1}\delta_{p_2 p'_2}...\delta_{p_D p'_D}.
\label{23}
\end{equation}

ii) the Lie derivative corresponding to rigid translations, which reduces
to a simple directional derivative on any form
\begin{equation}
{\cal L}_{c^i \partial_{z^i}} =[\partial,
c^i\partial_{dz^i}]_+=c^i \partial_{z^i},
\qquad c^i=const.  \in {\bf C}.
\label{24}
\end{equation}
The eigenvalue problem reads
\begin{equation}
c^i \partial_{z^i}\kappa =\gamma  \kappa ,
\qquad \gamma =c^i \gamma_i \in {\bf C}.
\label{25}
\end{equation}
and its solution is
\begin{equation}
\kappa_{\gamma}^{(p)}=e^{-{1 \over 2} \gamma^*{}^i \gamma_i}
 e^{\gamma_i z^i} (dz^1)^{p_1}(dz^2)^{p_2}...
(dz^D)^{p_D},
\label{26}
\end{equation}
after normalization, using
$\exp(\gamma^*{}^i \partial_{z^i})\Psi(z^j,dz^k)
=\Psi(z^j +\gamma^*{}^j,dz^k)$.
Coherent states of different
form degree are orthogonal, but the scalar
product of two arbitrary  coherent states is
\begin{equation}
\langle \kappa^{(p)}_{\gamma},\kappa^{(p')}_{\gamma'}\rangle
=e^{-{1 \over 2}(\gamma^*{}^i \gamma_i +\gamma^*{}'^i \gamma'_i
-2\gamma^*{}^i \gamma'_i)^2}
\delta_{p_1 p'_1}\delta_{p_2 p'_2}...\delta_{p_D p'_D}.
\label{27}
\end{equation}
The $\kappa^{(p)}_{\gamma}$ generalize the  coherent
states of the bosonic harmonic oscillator, which are contained for the
special case $p=0$. The characteristic properties of the
bosonic coherent states are preserved for arbitrary $p$.
In particular, they constitute an overcomplete set, if the
$\gamma_i$ are not restricted to a subset of the ${\bf C}^D$ plane,
 which is just complete \cite{per} and
   they are minimum uncertainty states with
respect to the position operators
$(1 / \sqrt{2})(z^i+
\partial_{z^i})$
and momentum operators
 $(i / \sqrt{2})(z^i-
\partial_{z^i})$
for any
$p$.
The coherent states above have to be distinguished from
''supercoherent states'' as discussed in \cite{fat}, which
employ Grassmann parameters.

\section{Concluding remarks}

The whole formalism is invariant under $U(D)$ transformations
$z^i \to   z'^i = \Lambda^i{}_j z^j$ where $\Lambda^i{}_j \in U(D)$,
which
is the symmetry group of the classical bosonic oscillator.
In fact, as is discussed in \cite{rit},
the full symmetry group of the SUSY oscillator
is the supergroup $U(D/D)$, which combines
the $U(D)$ transformation with the interchange of objects paired by
the SUSY charges.
The Hamiltonian generates
an abelian subgroup corresponding to evolution
$U(D/D)=U(1)(t) \times SU(D/D)$.
The strictly real
approach of \cite{thi} by contrast has only $O(D)$ in the bosonic
sector being promoted to $O(D/D)$, which contains
no continous abelian subgroup that  could account for evolution.

So far, we neither used any non-holomorphic quantities, nor did we use
a metric on the manifold.
By means of these additional ingredients, it is possible
 to give an integral version of the scalar product
eq.(\ref{11}) for a $p$- and a $q$-form by using the Hodge star
\begin{equation}
\langle  \Psi^{(p)}(z^i,dz^j)| \Xi^{(q)}(z^k,dz^l)\rangle
= {1 \over 2^p \pi^D} \int_{{\bf C}^D}e^{-z^m\bar{z}_m}
\Psi^{(p)}(z^i,dz^j)* \Xi^{(q)}(\bar{z}^k,d\bar{z}^l),
\label{29}
\end{equation}
where the Hodge star
is with respect to the standard euclidian metric
and in our coordinates the corresponding orthonormal frame is
$(\partial_{x^1},\partial_{y^1},\partial_{x^2},\partial_{y^2},$
$...,
 \partial_{x^D},\partial_{y^D})$.
Absorbing the exponential into the entries of the scalar product,
we have
\begin{equation}
\langle  \Psi^{(p)}|\Xi^{(q)}\rangle
={1 \over 2^p \pi^D} \int_{{\bf C}^D} \langle
\Psi^{(p)}|\bar{z}^i,d\bar{z}^j\rangle * \langle  \bar{z}^k,d\bar{z}^l
|\Xi^{(q)}\rangle.
\label{30}
\end{equation}
The scalar product vanishes whenever $p\ne q$, because for $p-q >0$
a ($p+2D-q$)-form vanishes by antisymmetry and for  $p-q <0$
we integrate over a set of measure zero in $2D$-dimensional space.
For  arbitrary elements of the exterior algebra,
eqns.(\ref{29}) or (\ref{30}) have to be applied after decomposition
into homogeneous $p$-forms.
For 0-forms the above version for the scalar product
reproduces the usual Bargmann-Fock prescription \cite{foc,bar}
and coincides with
the coherent state representation for the bosonic harmonic oscillator
 \cite{per}.
It is remarkable that we manage
to integrate fermionic quantities
with conventional integration, such that our integral version
of the scalar product is different from the usual Grassmann integration
\`a la Berezin \cite{be1}.

Finally, if we had developed the formalism on the entire ${\bf C}^D$,
thus considering
forms $\Gamma(z^i,dz^j,\bar{z}^k,d\bar{z}^l)$, we would have
needed a prepscription in order to select the  holomorphic forms that
represent the physical states.
This is accomplished by imposing $\bar{\partial}\Gamma=0$, while $\Gamma \ne
 \bar{\partial}...$.
 Therefore
$\Lambda {\bf C}^D_{\rm h}
=\bigoplus_{p=0}^D \Lambda^p {\bf C}^D_{\rm h} \equiv {\cal H}(
{\bf C}^D,\bar{\partial})
=\bigoplus_{p=0}^D {\cal H}^p({\bf C}^D,\bar{\partial})$
such that we are
actually working in the non-trivial cohomology sector of $\bar{\partial}$.

\vskip 0.5truecm
\noindent {\bf Acknowledgment:}

\noindent I am indebted to H. D. Dahmen and
D. Schiller for
helpful discussions.


\newpage

\begin{thebibliography}{99}

\bibitem{foc}V.A. Fock, Z. Phys. {\bf 49} (1928), 339.

\bibitem{bar}V. Bargmann, Commun. Pure and  Appl. Math.  {\bf 14} (1961),
187.

\bibitem{bar1} V. Bargmann, Rev. Mod. Phys. {\bf 34} (1962) 829.

\bibitem{nic}H. Nicolai, J. Phys. {\bf A9} (1976), 1497.

\bibitem{wit}E. Witten, Nucl. Phys. {\bf B185}, 513 (1981).

\bibitem{wit1}E. Witten, J. Diff. Geom., {\bf 17}, (1982), 661-692.

\bibitem{rit}M. de Crombrugghe, V. Rittenberg, Ann. Phys. {\bf 151}
(1983), 99.

\bibitem{gen}Gendenstein L.E., Kriv\'e I.V., Soviet Physics Uspekhi
{\bf 28}:8, 645-660 (1985).

\bibitem{cho}Y. Choquet-Bruhat, C. DeWitt-Morette, M. Dillard-Bleick:
{\em Analysis}, {\em Manifolds and Physics}, revised edition (North-Holland,
Amsterdam 1982).

\bibitem{wel}R.O. Wells: {\em Differential Analysis on Complex Manifolds}
 (Prentice-Hall, Inc., Englewood Cliffs, New Jersey 1973)

\bibitem{thi} H.-P. Thienel, J. Math. Phys. {\bf 36} (1995) 1192.

\bibitem{per}A. Perelomov: {\em Generalized Coherent States and Their
Applications} (Springer-Verlag, Berlin Heidelberg 1986).

\bibitem{fat}B. W. Fatyga, V. A. Kostelecky, M. M. Nieto,
D. R. Truax,  Phys. Rev.  {\bf D 43}, 1403 (1991).

\bibitem{be1}F.A. Berezin: {\em The Method of Second Quantization}
(Academic Press, New York 1966).

\end{thebibliography}
\end{document}